\documentclass[12pt,a4paper,notitlepage]{article}
\usepackage{amssymb,amsmath}
\usepackage[dvips]{graphicx}
\newcommand{\bsl}{\boldsymbol}

\newcommand{\bc}{\begin{center}}
\newcommand{\ec}{\end{center}}
\newcommand{\be}{\begin{equation}}
\newcommand{\ee}{\end{equation}}
\newcommand{\ba}{\begin{array}}
\newcommand{\ea}{\end{array}}
\newcommand{\bea}{\begin{eqnarray}}
\newcommand{\eea}{\end{eqnarray}}
\newcommand{\bt}{\begin{tabular}}
\newcommand{\et}{\end{tabular}}

\newcommand{\ov}{\overline}
\begin{document}
\begin{center}
{\Large\bf Heavy quark bound states above deconfinement} 

\vspace*{6mm}
{I. M. Narodetskii, Yu. A Simonov, A. I. Veselov}\\[5mm]
 Institute of Theoretical and Experimental Physics, Moscow
117218, Russia\\[2mm]
\end{center}

\vspace*{6mm}

\begin{abstract}
A comprehensive study of the color singlet heavy quark states
$J/\psi$, $\Upsilon$ and $\Omega_{bbb}$ above $T_c$ is given. We
use the Field Correlator Method (FCM) for nonperturbative $Q{\bar
Q}$ potentials and the screened Coulomb potential with the
$T$-dependent Debye mass computed up to two loops in the decon¯ned
phase of QCD. We calculate binding energies and melting
temperatures of heavy mesons and baryons in the deconfined phase
of quark-gluon plasma and the $J/\psi$ and $\Upsilon$
disintegration cross sections via the gluon absorption.

\end{abstract}
\vspace{5mm}


\section{Introduction}

 Early
applications  of in-medium heavy-quark potentials have employed a
phenomenological ansatz to implement color-screening effects into
the one-gluon-exchange (Coulomb) potential \cite{KMS:1988}. The
temperature dependence of this potential is encoded in the Debye
mass $m_D(T)$. If Debye screening of the Coulomb potential above
the temperature of deconfinement $T_c$ is strong enough, then
$J/\psi$ production in A+A collisions will be suppressed.
Therefore the 'gold-plated' signature of deconfinement in the
Quark-Gluon Plasma (QGP) was thought to be the $J/\psi$
suppression \cite{MS:1986}. Indeed, applying the Bargmann
condition \cite{bargmann} for the screened Coulomb potential at
$T=T_c$
\begin{equation}\label{eq:coulomb}
V_{\rm C}(r)=-\frac{4}{3}\cdot\frac{\alpha_s(r)}{r} \cdot
e^{-m_D(T_c)\,r},\end{equation}   we obtain the simple estimate
for the number of, say, $c{\overline c}$ $S$-wave bound states \be
n\leq \mu_c\int\limits_0^{\infty}|V_{\rm
C}(r)|r\,dr\,=\,\frac{4\alpha_s}{3}\cdot\frac{\mu_c}{m_D(T_c)}\,
,\label{eq:bargmann}\end{equation} where $\mu_c$ is the
constituent mass of the $c$-quark and for the moment we neglect
the $r$-dependence of $\alpha_s$. Taking $\mu_c=1.4$ GeV and
$\alpha_s=0.39$, we conclude that if $m_D(T_c)\geq 0.7$ GeV, there
is no $J/\psi$ bound state. Parenthetically, we note that for the
potential (\ref{eq:coulomb}) no light or strange mesons ($\mu\sim
300-500$ MeV) survive
. But this is not the full story.

There is a significant change of views on physical properties and
underlying dynamics of quark--gluon plasma (QGP), produced at
RHIC, see, {\it e.g.}, \cite{T:2009} and references therein.
Instead of behaving like a gas of free quasiparticles~--~quarks
and gluons, the matter created in RHIC interacts much more
strongly than originally expected.
It is more appropriate to describe the
nonperturbative (NP) properties of the QCD phase close to $T_c$ in
terms of the NP part of the QCD force rather than a strongly
coupled Coulomb force.

In the QCD vacuum, the NP quark--antiquark potential is {$V=\sigma
r$.} At $T \geq T_c$, $\sigma=0$, but that does not mean that the
NP potential disappears. In a recent paper \cite{NSV:2009} we
calculated binding energies for the lowest $Q\overline{Q}$ and
$QQQ$ eigenstates ($Q=c,b$) above $T_c$ using the NP
$Q\overline{Q}$ potential derived in the Field Correlator Method
(FCM) \cite{NST:2009} and the screened Coulomb potential.
The Debye radii were calculated for pure gluodynamics using Eqs.
(28),(29) of Ref. \cite{A:2003} and the parameters given there in.
In the present paper we refine the results of \cite{NSV:2009} and,
in particular, extend our analysis to the case of the running
$\alpha_s(r)$ \cite{conf:tallahassee}. We also calculate the
dissociation cross sections of $J/\psi$ and $\Upsilon$ in
collisions with gluons. Here as in Ref. \cite{NSV:2009} the Debye
mass is evaluated in quenched QCD (the number of light flavors
$n_f = 0$, $T_c=275$ MeV), the similar results using the Debye
mass evaluated for two light flavors ($T_c=165$ MeV) are presented
in Ref. \cite{NSV:2010}.\vspace{4mm}

\section{Field correlator method as applied to finite temperatures}

The NP quark-antiquark potential can be studied through the
modification of the correlator functions, which define the
quadratic field correlators of the nonperturbative vaccuum fields:
\begin{equation}<{\rm tr}\,F_{\mu\nu}(x)\Phi(x,0)F_{\lambda\sigma}(0)>\,=\,{\cal
A}_{\mu\nu;\lambda\sigma}\,D(x)+{\cal
B}_{\mu\nu;\lambda\sigma}\,D_1(x),\nonumber\end{equation} where
${\cal A}_{\mu\nu;\lambda\sigma}$ and ${\cal
B}_{\mu\nu;\lambda\sigma}$ are the two covariant tensors
constructed from  $g_{\mu\nu}$ and $x_{\mu}x_{\nu}$
\cite{NST:2009}, $\Phi(x,0)$ is the Schwinger parallel
transporter, $x$ is Euclidian. At $T\geq\,T_c$, one should
distinguish  the color electric correlators
$D^E(x),\,\,\,D^E_1(x)$ and color magnetic correlators
$D^H(x),\,\,\,D^H_1(x)$. Above $T_c$, the color electric
correlator $D^E(x)$ that defines the string tension at $T=0$
becomes zero \cite{Si:1991} and, correspondingly, $\sigma^E=0$.
The color magnetic correlators $D^H(x)$ and $D^H_1(x)$ do not
produce static quark--antiquark potentials, they only define the
spatial string tension $\sigma_s=\sigma^H$ and the Debye mass
$m_d\propto\sqrt{\sigma_s}$ that grows with $T$.

The main source of the NP static $Q\overline Q$ potential at
$T\,\geq\,T_c$ originates from the color--electric correlator
function $D^{E}_1(x)$:\begin{equation}\label{eq:potential} V_{\rm
np}(r,T)\,=\,\int\limits_0^{1/T}d\nu(1-\nu T)\int\limits_0^r
\lambda d\lambda\, D_1^{E}(x).\end{equation} In the confinement
region the function $D_1^{E}(x)$ was calculated in \cite{Si:2005}:
\begin{equation}D^{E}_1(x)\,=\,{B}\,\,\frac{\exp(-M_0\,x)}{x},\end{equation}
where $ B=6\alpha_s^f\sigma_fM_0$, $\alpha_s^f$ being the freezing
value of the strong coupling constant to be specified later,
$\sigma_f$ is the string tension at $T=0$, and the parameter $M_0$
has the meaning of the gluelump mass. In what follows we take
$\sigma_f=0.18$ GeV$^2$ and $M_0=1$ GeV. Above $T_c$ the
analytical form of $D_1^E$ should stay unchanged at least up to
$T\sim 2\,T_c$. The only change is $B\to B(T)=\xi(T)B$, where the
$T$-dependent factor
\be\xi(T)\,=\,\left(1-0.36\,\frac{M_0}{B}\,\frac{T-T_c}{T_c}\right)\ee
is determined by lattice data \cite{DMSV:2007}.
Integrating Eq. (\ref{eq:potential}) over $\lambda$, one
obtains\begin{equation}\label{eq:vnp} V_{np}(r,T)=\frac{B(T)}{M_0}
\int\limits_0^{1/T}\, (1-\nu T)\left(e^{-\nu
M_0}-e^{-\sqrt{\nu^2+r^2}\,M_0}\right)d\nu\,=\,V(\infty,T)-V(r,T),\end{equation}
where \be\label{eq:vinf}
V(\infty,T)=\frac{B(T)}{M_0^2}\left[1-\frac{T}{M_0}\left(1-\exp\left(-\frac{M_0}{T}\right)\right)\right],\ee
and \be\label{eq:integral} V(r,T)=\frac{B(T)}{M_0}
\int\limits_0^{1/T}\, (1-\nu
T)\exp(-\sqrt{\nu^2+r^2}\,M_0)d\nu.\ee The approximate expression
\be\label{eq:approximation} V(r,T)\approx
\frac{B(T)}{M_0^2}\left(K_1(x)x-\frac{T}{M_0}\exp(-x)(1+x)\right),\ee
where $x=M_0r$ and $K_1(x)$ is the McDonnald function, has been
used in \cite{NSV:2009}. For $T=T_c$ expressions
(\ref{eq:integral})and (\ref{eq:approximation}) are almost
indistinguishable. At $T\,>\,T_c$ the exact potential is slightly
more attractive, the difference between (\ref{eq:integral}) and
(\ref{eq:approximation}) increases with $T$. Even more drastic
approximation $V(r,T)\propto xK_1(x)$ was proposed \cite{Si:2005}.
This approximation was used in \cite{NSV:2009} to calculate the
$QQQ$ states.

\section{Coulomb potential}

We use the perturbative screened Coulomb potential
(\ref{eq:coulomb}) with the $r$-dependent QCD coupling constant
$\alpha_s(r,T)$. Note that in the entire regime of distances, for
which at $T=0$ the heavy quark potential can be described well by
QCD perturbation theory, $\alpha_s(r,T)$ remains unaffected by
temperature effects at least up to $T\le 3\,T_c$ and agrees with
the zero temperature running coupling $\alpha_s(r,0)=\alpha_s(r)$.
For our purposes, we find it convenient to define the
$r$--dependent coupling constant in terms of the ${\textbf q}^2$--
dependent constant $\alpha_B({\textbf q}^2)$ calculated in the
background perturbation theory (BPTh) \cite{Si:1995}:

\begin{equation}
\label{eq:alpha}
\alpha_s(r)\,=\,\frac{2}{\pi}\,\int\limits_0^{\infty}dq\,\,\frac{\sin\,qr}{q}\,\,\alpha_B({\textbf
q}^2).
\end{equation}
The formula for $\alpha_B({\textbf q}^2)$ is obtained by solving
the two-loop renormalization group equation for the running
coupling constant in QCD
\begin{equation}
\label{eq:alpha_V}
 \alpha_B({\textbf
q}^2)\,=\,\frac{4\pi}{\beta_0\,t}\left(1\,-\,\frac{\beta_1}{\beta_0\,^2}\,\frac{\ln
t}{t}\right),\,\,\,\,\,t\,=\,\ln\,\frac{{\textbf
q}^2+m_B^2}{\Lambda_V^2},
\end{equation}
where $\beta_i$ are the coefficients of the QCD $\beta$-function.
The parameter $m_B\sim 1$ GeV has the meaning of the mass of the
lowest hybrid excitation. The result can be viewed as arising from
the interaction of a gluon with background vacuum fields. Note
that $\alpha_B(r)$ increases with $\Lambda_V$ and, for fixed
$\Lambda_V$, decreases with $m_B$.

We employ the values $\Lambda_V\,=\,0.36\,{\rm GeV},\,\,\,
m_B\,=\,0.95\,{\rm GeV}$, which lie within the range determined in
\cite{BK2001}. The result is consistent with the freezing of
$\alpha_B(r)$ with a magnitude 0.563 (see Table 1 of
\cite{KNV:2009}
). The zero temperature
potential with the above choice of the parameters gives a fairly
good description of the quarkonium spectrum.

The Debye mass $m_D(T)$ in Eq. (\ref{eq:coulomb}) is expressed in
terms of the spatial string tension $\sigma_s(T)$ due to
chromomagnetic confinement: $ m_D(T)=2.06\sqrt{\sigma_s(T)}$. The
latter has been computed nonperturbatively up to two loops in the
deconfined phase of QCD \cite{A:2003}. As was stated above, in
this paper, we consider the pure-gauge {\it SU}(3) theory
($T_c=275$ MeV), for which $m_D$ varies between 0.8 GeV and 1.4
GeV, when $T$ varies between $T_c$ and $2\,T_c$.

\section{Results}
\subsection{Quark-antiquark states}
 In the framework of the FCM, the masses of heavy
quarkonia are defined as \begin{equation} \label{eq:mass}M_{Q\bar
Q}\,=\,\frac{m_Q^2}{\mu_{Q}}\,+\,\mu_Q\,+\,E_0(m_Q,\mu_Q),\end{equation}
$E_0(m_Q,\mu_Q)$ is an eigenvalue of the Hamiltonian
\be\label{eq:H} H=H_0+V_{\rm np}+V_{\rm C},\ee $m_Q$ are the bare
quark masses, and einbeins $\mu_i$ are treated as c-number
variational parameters. The eigenvalues $E_0(m_i,\mu_i)$ of the
Hamiltonian (\ref{eq:H}) are found as functions of the bare quark
masses $m_i$ and einbeins $\mu_i$, and are finally minimized with
respect to the $\mu_i$. With such simplifying assumptions the
spinless Hamiltonian $H_0$ takes an apparently nonrelativistic
form, with einbein fields playing the role of the constituent
masses of the quarks. Once $m_Q$ is fixed, the quarkonia spectrum
is described.
The dissociation points are defined as those temperature values
for which the energy gap between $V(\infty,T)$ and $E_0(T)$
disappears.

In our calculations, we use the quark--antiquark potentials whose
parameters are listed in Table \ref{tab:parameters}. The potential
I was employed in \cite{NSV:2009}. This potential uses the
approximation (\ref{eq:approximation}) for $V(r)$ and the constant
value $\alpha_s=0.35$ for the Coulomb potential. The potential II
is the same potential but with the running $\alpha_s(r)$ given by
Eq. (\ref{eq:alpha}). In this case, we have slightly changed the
parameter $M_0$ to preserve the value of $V(\infty,T_c)=0.508$ GeV
that agrees with lattice estimate for the free quark--antiquark
energy~\footnote{~However, the difference in the parameter $M_0$
causes the small difference of $V(\infty,T)$ for $T\,>\,T_c$, see
tables \ref{tab:cc}, \ref{tab:bb}.}. Note that for the potential
in Eq. (\ref{eq:approximation}) the Bargmann integral
(\ref{eq:bargmann}) is
\begin{equation}\frac{\mu_QB}{M_0^4}\,(2-3\frac{T}{M_0})\end{equation}The potentials III and IV are
defined by  the exact integral representation (\ref{eq:integral})
for $V(r)$ and correspond to the constant and running  $\alpha_s$,
respectively.

\begin{figure}
\begin{center}
\includegraphics[width=100mm, keepaspectratio=true,angle=-90]{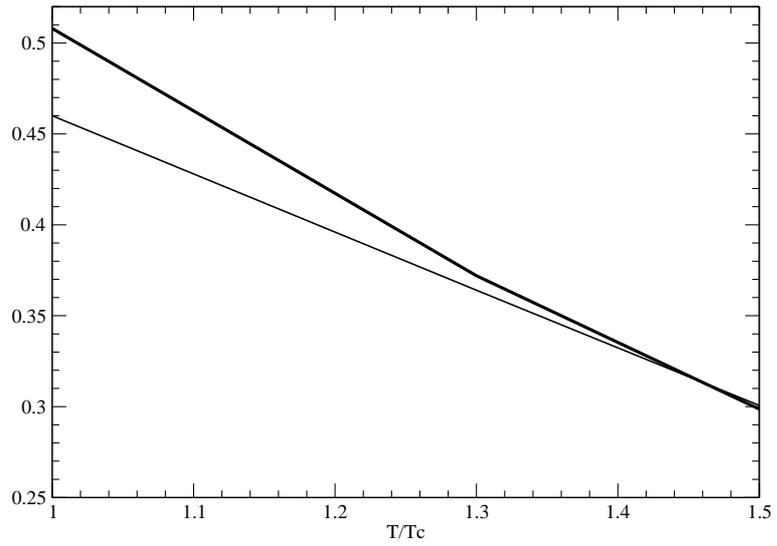}
\vskip -2cm
\end{center}
\caption{$V(\infty,T)$ (thick curve) and $E_0(T)$ (thin
 curve) are plotted against $T/T_c$
 for the $J/\psi$ (the potential IV). $V(\infty,T)$ and $E_0(T)$ are given in units of GeV.
}
\label{fig:psi}
\end{figure}

\begin{figure}
\begin{center}
\includegraphics[width=120mm, keepaspectratio=true,angle=-90]{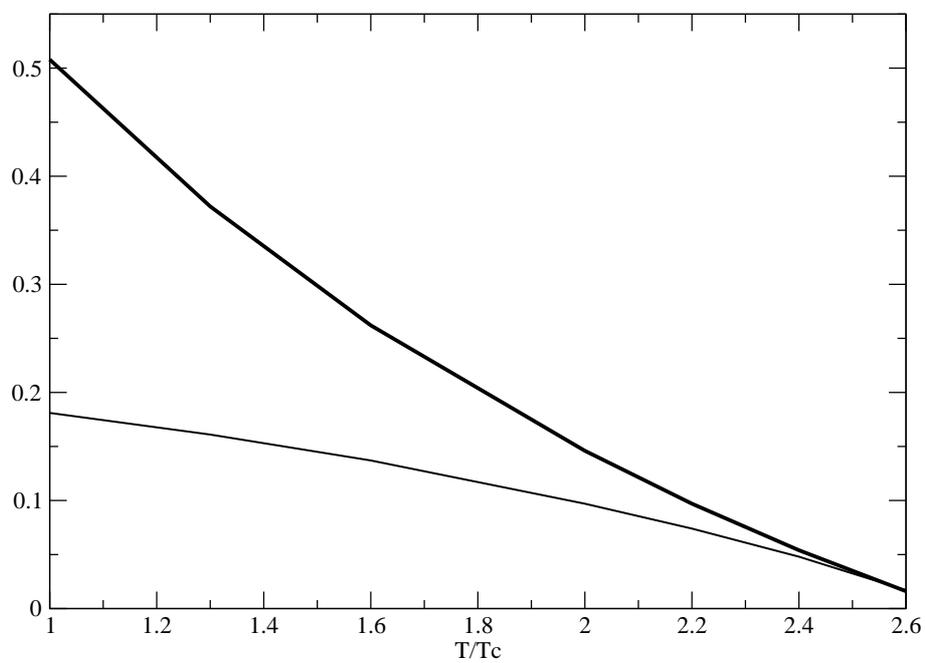}
\end{center}
\caption{The same as in Fig. \ref{fig:psi} for the ground
$\Upsilon$ state} \label{fig:upsilon}
\end{figure}

\begin{figure}
\begin{center}
\includegraphics[width=120mm, keepaspectratio=true,angle=-90]{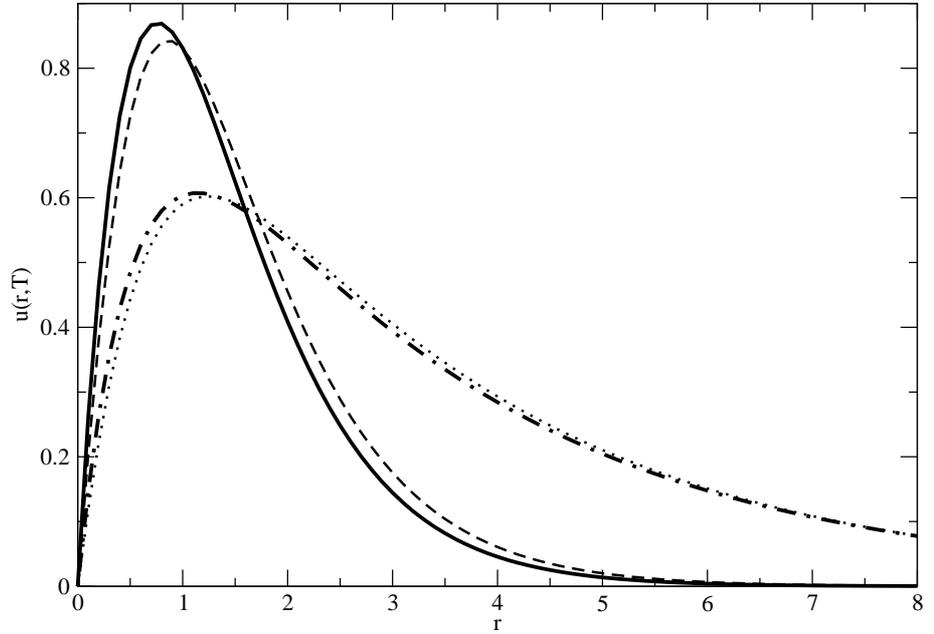}
\end{center}
\caption{The ${\ov b}b$ wave function $u(r,T)$ normalized to unity
($r$ in units of GeV$^{-1}$). Solid
 and dashed curves show the $u(r,T_c)$ for the potential I and II, respectively, dot-dashed  and dot curves
 show the $u(r,2T_c)$ for the potentials I and II.}
\label{fig:wavefunctions}
\end{figure}

We display in Figs. \ref{fig:psi}, \ref{fig:upsilon} the binding
energies of the 1S $J/\psi$ and $\Upsilon$ mesons above the
deconfinement temperature.
The details of the calculation are presented in Tables
\ref{tab:cc}, \ref{tab:bb}~\footnote{~In tables \ref{tab:cc},
\ref{tab:bb} the results for the potential I that were previously
reported in \cite{NSV:2009} are quoted for comparison.}. In these
tables we present the constituent quark masses $\mu_Q$ for
$c\overline{c}$ and $b\overline{b}$, the binding energies
$E_0-V_{np}(\infty,T)$, the mean squared radii $r_0=\sqrt{<r^2>}$,
and the masses of the $Q\overline{Q}$ mesons. We employ $m_c=1.4$
GeV and $m_b=4.8$ GeV. As in the confinement region, the
constituent masses $\mu_Q$ only slightly exceed the bare quark
masses $m_Q$ that reflect smallness of the kinetic energies of
heavy quarks. We also mention that the account of the running
$\alpha_s(r)$ in the Coulomb potential produces a tiny effect as
compared with the case of a constant $\alpha_s=0.35$ both for the
energies (compare lines I and II and III and IV in Tables
\ref{tab:cc}, \ref{tab:bb}) and $b{\bar b}$ wave functions in Fig.
\ref{fig:wavefunctions}. Both for the charmonium and for the
bottomonium states the energy gap $V(\infty,T)-E_0(T)$  gets
smaller and the mean square radius $r_0$ gets larger as the
temperature grows. We find no excited states 1P and 2S states,
although the unbound 2S $b{\bar b}$ state appears to be very close
to the threshold.

At $T=T_c$ we obtain the weakly bound $c{\overline c}$ state. The
potential II predicts small additional binding $\sim$ 20 MeV as
compared with the binding for potential I.
At $T=1.3\,T_c$ the difference of $V(\infty,T)-E_0(T)$ calculated
for different potentials comprises only a few MeV. The melting
temperature for the case IV is $\sim\,1.3\,T_c$. The charmonium
masses lie in the interval 3.1--3.3 GeV, that agrees with the
results of \cite{DMSV:2007}. Note that immediately above $T_c$ the
mass of the  $c\overline{c}$ state is about 0.2 GeV higher than
that of $J/\psi$.

As expected, the $\Upsilon$ state is much more bound and remains
intact up to the larger temperatures, $T\,\sim\, 2.3\,T_c$ (all
the details of calculation can be inferred from Table
\ref{tab:bb}). The masses of the {\it L} = 0 bottomonium lie in
the interval 9.7--9.8 GeV, about 0.2--0.3 GeV higher than 9.460
GeV, the mass of $\Upsilon(1S)$ at $T=0$. At $T=T_c$ the
$b{\overline b}$ separation $r_0$ is 0.25 fm, which is compatible
with $r_0=0.28$ fm at $T=0$. At the melting point $r_0\to\infty$.
Note that the 1S bottomonium undergoes very little modification
till
 $T\,\sim 2\,T_c$. The results agree with those found previously for a constant
$\alpha_s=0.35$ \cite{NSV:2009}. The melting temperatures for the
$J/\psi$ and $\Upsilon$ are shown in Table \ref{tab:melting}. The
results for the $\Upsilon$ are in agreement with the lattice study
of Ref. \cite{aarts:2010}. Note that in our calculations we
neglect the spin-spin force, therefore the $J/\psi$ and $\eta_c$
mesons appear degenerate, as well as the $\Upsilon$ and $\eta_b$.
This degeneracy is expected to be removed by a short range
spin-spin interaction, whose effect, at T = 0, is often treated
perturbatively assuming a contact interaction.
\subsection{Dissociation of $J/\psi$ and $\Upsilon$ in collision with gluons}

Heavy quark bound states are important probes of the dynamics in
the QGP. Charmonium suppression has been observed at a variety of
energies at SPS \cite{NA60:2009} and RHIC \cite{PHENIX:2008}.
While the melting of bound states certainly reduces quarkonium
production, the converse is not true: different, even competing
effects make it difficult to interpret charmonium suppression
patterns. It has been noted that such effects should be less
significant for bottomonium \cite{Rapp:2008}.

Previous treatment of the dissociation of heavy quarkonium by the
absorption of a gluon was carried out \cite{P:1979} using the
operator product expansion in the the large $N_c$ limit and
hydrogen states to evaluate the transition matrix elements. In
heavy quarkonia of interest, the radial dependence of the
quark-antiquark potential and the corresponding wave functions
differs from the Coulomb potential. The calculation of the
dissociation cross section for a color $E1$ transition can be well
described by the potential model, following the results of
Akhiezer and Pomeranchuk \cite{AP:1948} and Blatt and Weisskopf
\cite{BW:1952} obtained for the photo-disintegration of a
deuteron. At low energies, the dominant dissociation cross section
is the $E1$ color-electric dipole transition for which the quark
final state will be the continuum $(Q{\bar Q})$ 1P state. We
calculate the cross section for the quarkonium dissociation after
a gluon impact similarly the calculation of the deuteron
disintegration via the photon absorption as has been done in Refs.
\cite{Blaschke:2005},
\cite{Wong:2005}.

\begin{figure}
\begin{center}
\includegraphics[width=120mm, keepaspectratio=true]{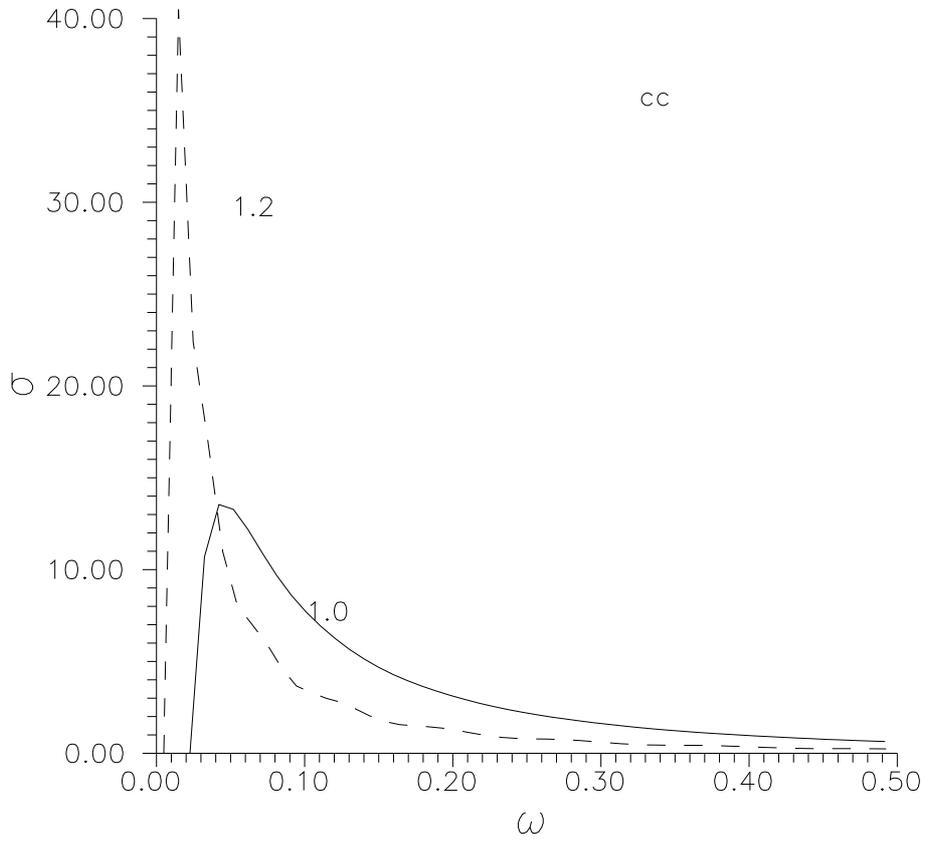}

\caption{Cross sections $\sigma$ for $J/\psi$  dissociation in the
potential model vs gluon energy $\omega$. The curves are labelled
by the ratios $T/T_c$. $\sigma$ and $\omega$ are given in units
GeV$^{-2}$ and GeV, respectively.
 } \label{fig:dissociation_cc}
\end{center}
\end{figure}

\begin{figure}
\begin{center}
\includegraphics[width=120mm, keepaspectratio=true]{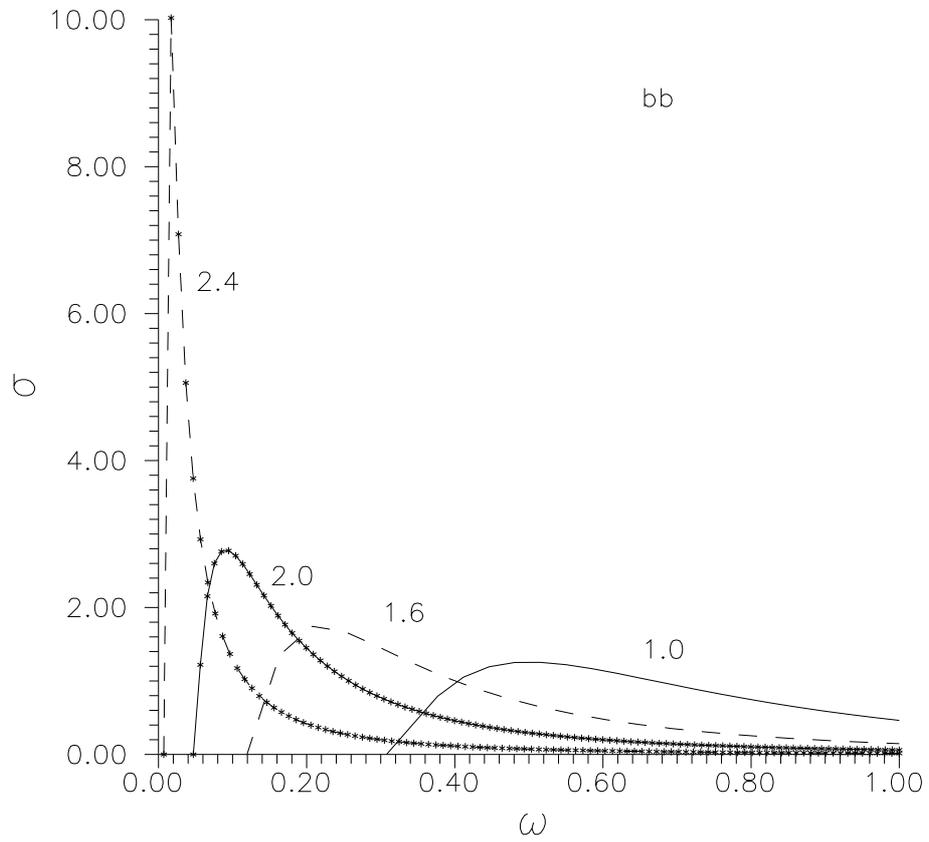}

\caption{Cross sections for $b{\bar b}$ dissociation in the
potential model. The notations are the same as in Fig.
\ref{fig:dissociation_cc} } \label{fig:dissociation_bb}
\end{center}
\end{figure}

An initial bound $(Q{\bar Q})_{1S}$ state with a binding energy
$\varepsilon(T)=V(\infty,T)-E_0(T)$ relative to the threshold in
the color-singlet potential~\footnote{~$\varepsilon(T)\geq\,0$
below the melting point.} absorbs a $E1$ gluon of energy $\omega$,
and is excited to the color-octet final state $(Q{\bar Q})_{1P}$
with the energy
\begin{equation}\frac{k^2}{\mu_Q}=\omega-\varepsilon(T).\label{eq:energy}\end{equation}
The corresponding cross section reads
\begin{equation} \sigma_{(Q{\bar
Q)g}}(\omega)=\frac{4\pi\alpha_{gQ}}{3}\,\frac{k^2+k_0^2}{k}\left(\int\limits_0^{\infty}u_{1P}(r)u_{1S}(r,T)rdr\right)^2,
\end{equation}
where $u_{1S}(r,T)$ is the wave function of the $Q{\bar Q}$ bound
state at the temperature $T$ normalized as
\begin{equation}
\int\limits_0^{\infty}|u_{1S}(r,T)|^2dr\,=\,1,\end{equation}
$\alpha_{gQ}=\alpha_s/6$, $u_{1P}(r)$ is the free $P$-wave $Q{\bar
Q}$ wave function
\begin{equation}
u_{1P}(r)=\frac{\sin(kr)}{kr}-\cos(kr),
\end{equation}
$k$ is the momentum of outgoing quarks in Eq. (\ref{eq:energy}),
and
\begin{equation}k_0^2=2\mu_Q\,\varepsilon(T).\end{equation}
The results are shown in Figs \ref{fig:dissociation_cc},
\ref{fig:dissociation_bb}. They are in a qualitative agreement
with those of Ref. \cite{Blaschke:2005} where the $Q{\bar Q}$
potential was
identified with the color singlet heavy quark free energy above
$T_c$ taken from quenched lattice QCD simulations.

 \vspace{3mm}
 \subsection{$QQQ$ baryons at $T\geq T_c$}\label{section:QQQ}
\begin{figure}
\begin{center}
\includegraphics[width=120mm, keepaspectratio=true,angle=-90]{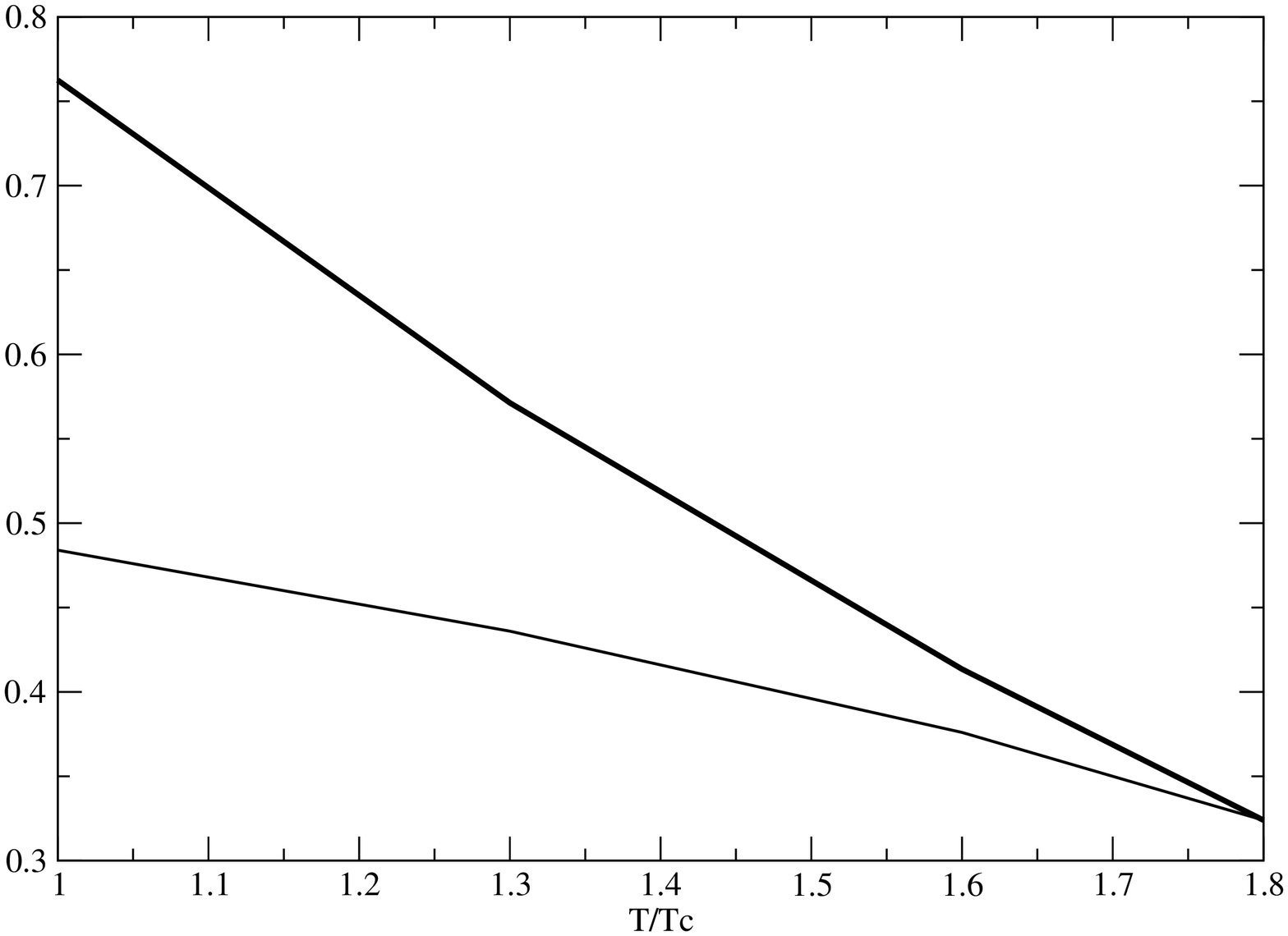}

\caption{ ${ \cal V}(\infty,T)$ (thick curve) and $E_0(T)$ (thin
 curve) in units of GeV
 for the $\Omega_{bbb}$ (the potential I)
} \label{fig:bbb}
\end{center}
\end{figure}

The three-quark potential is given by \be\label{eq:QQQ}
V_{QQQ}\,=\,\frac{1}{2}\,\sum_{i<j}\,V_{Q{\bar Q}}(r_{ij},T),\ee
where $\frac{1}{2}$ is the color factor. We solve the three-quark
Schr\"{o}dinger equation by the hyperspherical harmonics method.
Using the three-body Jacobi coordinates \be\label{eq:Jacobi}
{\bsl\rho}=\sqrt{\frac{\mu_{Q}}{2}}\,(\bsl{r}_1-\bsl{r}_2),\,\,\,\,\,\,\,\,\,\,\,
\bsl{\lambda}=\sqrt{\frac{2}{3}\,\mu_Q}
\left(\frac{\bsl{r}_1+\bsl{r}_2}{2}-\bsl{r}_3\right)\ee the wave
function $\psi(\bsl{\rho},\bsl{\lambda})$ in the hypercentral
approximation is written as
\be\Psi(\bsl{\rho},\bsl{\lambda},T)\,=\,\frac{1}{\sqrt{\pi^3}}\,\frac{u(R,T)}{R^{5/2}},
\ee
where the hyperradius \be\label{eq:R2}
R^2\,=\,\bsl{\rho}^2+\bsl{\lambda}^2\,=\,
\frac{\mu_Q}{3}\,\left(r_{12}^2\,+\,r_{23}^2\,+\,r_{31}^2\right)\ee
is invariant under quark permutations,
and\be\rho\,=\,R\,\sin\theta,\,\,\,\,\,
\lambda\,=\,R\,\cos\theta,\,\,\,\, 0 \le \theta \le \pi/2.\ee
Averaging the three-quark potential (\ref{eq:QQQ}) over the
six-dimensional sphere, one obtains the one-dimensional
Schr\"odinger equation for the reduced function $u(R,T)$:
\be\label{eq:se}\frac{d^2
u(R,T)}{dR^2}\,+\,2\left[E_0-\frac{15}{8\,R^2}-
\frac{3}{2}\left({\cal V}_C(R,T)+{\cal
V}(R,T)\right)\right]u(R,T)=0,\ee where \be {\cal V}_C(R,T)\,=\,
-\frac{4}{3}\,\alpha_s\,\int\limits_0^{\pi/2}
\exp(\,-\,m_d(T){\hat R})\,\frac{d\Omega_6}{\hat R},\ee`
\be\label{eq:v3Q} {\cal
V}(R,T)=V(\infty,T)-\frac{\xi(T)B}{M_0}\int\limits_0^{\pi/2}\left(K_1({\hat
R}){\hat R}-\frac{T}{M_0}\,e^{-{\hat R}}(1+{\hat
R})\right)d\Omega_6,\ee $V(\infty,T)$ being given by Eq.
(\ref{eq:vinf}), and   \be {\hat
R}=\frac{M_0R\sin\theta}{\mu_Q/2},\,\,\,\,\,\,
d\Omega_6=\frac{16}{\pi}\sin^2\theta\cos^2\theta\,d\theta.\ee
The temperature-dependent mass of the colorless $QQQ$ states is
defined as \be\label{eq:mQQQ} M_{QQ
Q}\,=\,\frac{3}{2}\frac{m_Q^2}{\mu_{Q}}\,+\,\frac{3}{2}\,\mu_Q\,+\,E_0(m_Q,\mu_Q),\ee
where $\mu_Q$ are now defined from the extremum condition imposed
on $M_{QQQ}$ in (\ref{eq:mQQQ}) \be\frac{\partial
M_{QQQ}}{\partial \mu_Q}=0\ee Note that the  average interquark
distances are \be\label{eq:rr}\sqrt{<r_{ij}^2>}=\sqrt{\frac{<
R^2>}{\mu_Q}},\ee see Eq. (\ref{eq:R2}).~ The bound $QQQ$ state
exists if $E_0(m_Q,\mu_Q)\,\leq\,{\cal V}_{QQQ}(\infty,T)$, where
\be\label{eq:VQQQinfinity}{\cal
V}_{QQQ}(\infty,T)\,=\,\frac{3}{2}\,V(\infty,T).\ee In our
three-quark calculations we use the potential I. For this
potential there is no bound $\Omega_{ccc}$ states~\footnote{The
same should be true for the potential III, because at $T=T_c$ both
potentials coincide. Moreover, since the effect of the running
$\alpha_s$ and the change of the parameter $M_0$ is almost
negligible, we conclude that $\Omega_{ccc}$ is unbound for all the
potentials listed in Table \ref{tab:parameters}.}. However, in all
our calculations the $\Omega_{ccc}$ was found to lie almost at
threshold. The $\Omega_{bbb}$ survives up to $T\sim 1.8\,T_c$, see
Table
 \ref{tab:bbb} and Fig. \ref{fig:bbb}~\footnote{The small difference between the results reported in
 Table \ref{tab:bbb} and those of Table 4 of \cite{NSV:2009} is due to the
 difference of the $r$-dependent part of the nonperturbative quark--antiquark potential,
 see footnote 1.}.
\section{Conclusions}

In conclusion, we have calculated binding energies and melting
temperatures for the lowest eigenstates in the $c\overline{c}$,
$b\overline{b}$, and $bbb$ systems. The color electric forces due
to the nonconfining correlator $ D^E _1$  survive in the
deconfined phase and they can support bound states at $T\,>\,T_c$.
For what concerns the charm states, we find that $J/\psi$ survive
up to $T\,\sim\,1.3\,T_c$, and there is no bound $\Omega_c$ state
at $T\geq\,T_c$. On the other hand, the $b\overline{b}$ and $bbb$
states survives up to higher temperature, $T\sim 2.6\,T_c$ and
$T\sim 1.8\,T_c$, respectively. This suggests that the systems are
strongly interacting above $T_c$.



This work was supported in part by RFBR Grant 09-02-00629.

\begin{table}
\caption{Parameters of the non-perturbative quark-antiquark
potentials described in the text. $B$ and $M_0$ are in units
GeV$^3$ and GeV, respectively. The potentials I-IV correspond to
the value of $V(\infty,T_c)=0.508$ GeV}\vspace{5mm}

\centering
\begin{tabular}{|c|c|c|c|c|c|c|}\hline\hline
&$V(r)$&$\alpha_s^f$&$\alpha_s(r)$&$B$&$M_0$\\\hline I&~~Eq.
(\ref{eq:approximation})&0.6&0.35&0.583&0.9\\\hline II&~~Eq.
(\ref{eq:approximation})&0.563&Eq.(\ref{eq:alpha})&0.494&0.813\\\hline
III&~~Eq. (\ref{eq:integral})&0.6&0.35&0.583&0.9\\\hline IV&~~Eq.
(\ref{eq:integral})&0.563&Eq.(\ref{eq:alpha})&0.494&0.813\\

\hline\hline
\end{tabular}

 \label{tab:parameters}\end{table}

  \begin{table}\caption{Details of the calculation of the $c\overline{c}$ states as a function of the
temperature above the deconfinement region.~$V(\infty,T)$,
$\mu_c$, $E_0-V_{np}(\infty)$, and $M_{c\overline{c}}$ are given
in units GeV, $r_0$ in units GeV$^{-1}$.~$m_c\,=\,1.4$
GeV}\centering \vspace{5mm}

\begin{tabular}{|c|c|c|c|c|c|c|}\hline\hline
$T/T_c$&Potential&$V(\infty,T)$&$\mu_c$&$E_0-V_{np}(\infty)$&$r_0$&$M_{c\overline{c}}$\\\hline
1~~&I&0.508&1.451&-\,\,0.019&7.53&3.291\\
&II&0.509&1.469&-\,\,0.040&6.07&3.271\\
&III&0.508&1.454&-\,\,0.022&7.24&3.288\\
&IV&0.509&1.473&-\,\,0.048&5.75&3.264\\\hline
1.3&I&0.381&1.419&+\,0.006&10.50&3.186\\
&II&0.372&1.425&+\,0.001&9.75&3.173\\
&III&0.381&1.424&+\,\,0.001&9.91&3.183\\
&IV&0.372&1.434&-\,\,0.008&8.72&3.165\\\hline
1.6&IV&0.262&1.416&+\,0.007&10.74&3.069\\\hline\hline
\end{tabular}
 \vspace{3mm}
\label{tab:cc}\end{table}

 \begin{table}\caption{Details of the calculation of the $b\overline{b}$ states as a function of the
temperature above the deconfinement region.~The notations are the
same as in Table \ref{tab:cc},~$m_b\,=\,4.8$ GeV}\centering
\vspace{5mm}

\begin{tabular}{|c|c|c|c|c|c|c|}\hline\hline
$T/T_c$&Potential&$V(\infty,T)$&$\mu_b$&$E_0-V_{np}(\infty)$&$r_0$&$M_{b\overline{b}}$\\\hline\hline
1~~&I&0.508&4.984&-\,\,0.300&1.27&9.815\\
&II&0.508&4.953&-\,\,0.315&1.32&9.798\\
&III&0.508&4.985&-\,\,0.308&1.26&9.807\\
&IV&0.509&4.953&-\,\,0.326&1.31&9.786\\\hline
1.3&I&0.381&4.950&-\,0.183&1.55&9.802\\
&II&0.372&4.925&-\,0.187&1.61&9.788\\
&III&0.381&4.953&-\,\,0.200&1.51&9.785\\
&IV&0.372&4.928&-\,\,0.211&1.55&9.764\\\hline
1.6&I&0.275&4.915&-\,0.095&2.06&9.783\\
&II&0.262&4.896&-\,0.093&2.14&9.772\\
&III&0.275&4.921&-\,0.119&1.91&9.759\\
&IV&0.262&4.902&-\,0.126&1.94&9.739\\\hline
2.0&I&0.162&4.863&-\,0.021&4.25&9.742\\
&II&0.146&4.851&-\,0.017&4.66&9.729\\
&III&0.162&4.878&-\,0.046&3.03&9.717\\
&IV&0.146&4.866&-\,0.048&3.05&9.698\\\hline
2.2&I&0.115&4.832&-\,0.003&7.62&9.712\\
&II&0.097&4.823&-\,0.001&8.51&9.696\\
&III&0.115&4.855&-\,0.023&4.32&9.693\\
&IV&0.097&4.847&-0.023&4.34&9.674\\\hline
2.3&I&0.093&~~4.818&~~+\,0.001~~&~~9.52~~&~~9.694\\
&II&0.075&4.813&+0.002&10.17&9.677\\\hline
2.4&III&0.073&4.831&-0.007&6.80&9.665\\
&IV&0.054&4.827&-0.007&6.91&9.648\\\hline
2.6&III&0.034&4.812&+0.001&9.91&9.635\\
&IV&0.016&4.809&+0.001&10.18&9.617\\
\hline\hline
\end{tabular}
 \vspace{3mm}

\label{tab:bb}\end{table}
 \begin{table}\caption{Dissociation temperatures (in units of $T_c$) for  $c\overline{c}$
 and $b\overline{b}$ states
}\centering
\vspace{5mm}

\begin{tabular}{|c|c|c|c|c|}\hline\hline
&I&II&III&IV\\\hline $J/\psi$&~~1.24&1.29&1.29&1.46\\\hline
$\Upsilon$&~~2.27&2.29&2.57&2.57\\

\hline\hline
\end{tabular}
 \vspace{3mm}
\label{tab:melting}\end{table}

\newpage
 \begin{table}\caption{Details of the calculation of $bbb$ baryon
 above the deconfinement
 region
 for the potential I of Table \ref{tab:parameters}. ${\cal
V}_{QQQ}(\infty,T)$ is defined by Eq.
(\ref{eq:VQQQinfinity}).~$\sqrt{\overline {R^2}}$ is related to
the interquark distances by Eq. (\ref{eq:rr}). Dimensions are the
same as in Table \ref{tab:cc}.~$m_b\,=\,4.8$ GeV}\centering
\vspace{15mm}

\begin{tabular}{|c|c|c|c|c|c|}\hline\hline
$\frac{T}{T_c}$&$\,\,\,{\cal
V}_{QQQ}(\infty,T)$&$\mu_b$&$E_0-{\cal V}(\infty,T)$
&$\sqrt{\overline {R^2}}$&$M_{bb{b}}$\\\hline\hline
1~~&0.763&4.943&-\,\,0.279&3.73&14.890\\\hline
1.3&0.571&4.906&-\,0.135&4.80&14.840\\
\hline
1.6&0.414&4.865&-\,0.037&7.17&14.777\\
\hline 1.8&0.324&4.835&+\,0.001&10.00&14.724\\\hline\hline
\end{tabular}
 \vspace{3mm}
\label{tab:bbb}\end{table} \vspace{40mm}


\vspace{5mm}

\vspace{5mm}



\begin{thebibliography}{99}
\bibitem{KMS:1988}F. Karsch, M.T. Mehr and H. Satz, Z. Phys. C {\bf 37}, 617 (1988).
\bibitem{MS:1986} T.~Matsui and H.~Satz, Phys. Lett. {B\bf 178},
416 (1986).
\bibitem{bargmann}V.~Bargmann, Proc. Nat. Acad. Sci. (USA), {\bf 38}, 961
(1952).
\bibitem{T:2009} M.~J.~Tannenbaum, Rep. Prog. Phys.
{\bf 69},  2005 (2006).
\bibitem{NSV:2009}I.~M.~Narodetskiy, Yu.~A.~Simonov,
A.~I.~Veselov, JETP Lett. {\bf 90}, 232 (2009) [Pisma Zh. Eksp.
Teor. Fiz. {\bf 90}, 254 (2009)].
\bibitem{NST:2009}
A.~V.~Nefediev, Yu.~A.~Simonov, M.~A.~Trusov, Int. J. Mod. Phys.
{\bf E18}, 549 (2009).
\bibitem{A:2003} N. O. Agasian, Phys. Lett. B{\bf 562},
257 (2003) arXiv:0303127 [hep-ph].
\bibitem{conf:tallahassee} I.~M.~Narodetskiy, Yu.~A.~Simonov,
A.~I.~Veselov, AIP Conf. Proc. 1257: 804-807, 2009

\bibitem{NSV:2010}I.~M.~Narodetskiy, Yu.~A.~Simonov,
A.~I.~Veselov, arXiv:1012.0890 [hep-ph].
\bibitem{Si:1991}Yu.~A.~Simonov, JETP Lett. {\bf 54}, 249 (1991), {\bf 55},  605 (1992);
Phys. Atom. Nucl. {\bf 58}, 309 (1995).
\bibitem{Si:2005} Yu.~A.~Simonov, Phys.~Lett. B {\bf 619}, 293
(2005), Phys. Atom. Nucl. {\bf 69}, 528 (2006).
\bibitem{DMSV:2007}
A.~DiGiacomo, E.~Meggiolaro, Yu.~A.~Simonov, and A.~I.~Veselov,
Phys.~Atom.~Nucl. {\bf 70}, 908 (2007).
\bibitem{Si:1995} Yu.~A.~Simonov, Phys. Atom. Nucl. {\textbf 58}, 107 (1995) [Yad. Fiz. {\textbf 58},
113 (1995)].





\bibitem{BK2001}
A.~M.~Badalian and D.~S.~Kuzmenko, Phys. Rev. D {\bf 65}, 016004
(2002).
\bibitem{KNV:2009} R.~Ya.~Kezerashvili, I.~M.~Narodetskiy, and A.~I.~Veselov, Phys.
Rev. D \textbf{79}, 034003 (2009).

\bibitem{aarts:2010} G. Aarts et al., arXiv: 1010.3725 [hep-lat].
\bibitem{NA60:2009} R. Arnaldi [NA60 Collaboration], Nucl. Phys. A{\bf 830},
345C (2009).
\bibitem{PHENIX:2008} A. Adare {\it et al.} [PHENIX Collaboration], Phys. Rev. Lett. {\bf
101},
 122301 (2008).\bibitem{Rapp:2008}  R. Rapp, D. Blaschke and P.
Crochet, arXiv:0807.2470 [hep-ph].
\bibitem{P:1979} M. E. Peskin, Nucl. Phys. B{\bf 156}, 365
(1979), G. Bhanot and M. E. Peskin, Nucl. Phys. B{\bf 156}, 391
(1979).
\bibitem{AP:1948} A.Akhieser and I.Pomeranchuk, {\it Some Problems
of Atomic Nucleus Theory}, Moscow-Leningrad, 1948.
\bibitem{BW:1952} J. M. Blatt and V. F. Weisskopf, Theoretical Nuclear
Physics, John Wiley and Sons, New York, 1952.
\bibitem{Blaschke:2005} D.~Blaschke, O.~Kaczmarek, E.~Laermann, V.~Yudichev, Eur. Phys.
J. C{\bf 43}, 81 (2005).
\bibitem{Wong:2005} Cheuk-Yin Wong, Phys. Rev. C{\bf 72}, 034906
(2005).
\end{thebibliography}
\end{document}